\documentclass[prb,aps,twocolumn,superscriptaddress,showpacs]{revtex4-1}

\usepackage{graphicx}
\usepackage{amsmath}
\usepackage{amsfonts}
\usepackage{amssymb} 

\newcommand{\lel}{l}
\newcommand{\rel}{r}
\newcommand{\bgel}{bg}
\newcommand{\chl}{L}
\newcommand{\bgd}{d}
\newcommand{\vds}{V}
\newcommand{\td}{\tilde}

\begin{document}

\title{Self heating and nonlinear current-voltage characteristics in bilayer graphene}

\author{J. K. Viljas}
\altaffiliation{Contributed equally to this work}
\affiliation{Low Temperature Laboratory, Aalto University, P.O.Box 15100, FI-00076 AALTO, Finland}
\author{A. Fay}
\altaffiliation{Contributed equally to this work}
\affiliation{Low Temperature Laboratory, Aalto University, P.O.Box 15100, FI-00076 AALTO, Finland}
\author{M. Wiesner}
\affiliation{Low Temperature Laboratory, Aalto University, P.O.Box 15100, FI-00076 AALTO, Finland}
\affiliation{Faculty of Physics, Adam Mickiewicz University, 61-614 Poznan, Poland}
\author{P. J. Hakonen}
\affiliation{Low Temperature Laboratory, Aalto University, P.O.Box 15100, FI-00076 AALTO, Finland}

\date{\today}

\pacs{73.63.-b, 73.23.-b, 73.50.Fq, 72.80.Vp}

\begin{abstract}
We demonstrate by experiments and numerical simulations that the
low-temperature current-voltage characteristics in diffusive bilayer
graphene (BLG) exhibit a strong superlinearity at finite bias
voltages. The superlinearity is weakly dependent on doping and on the
length of the graphene sample. This effect can be understood as a
result of Joule heating. It is stronger in BLG than in monolayer
graphene (MLG), since the conductivity of BLG is more sensitive
to temperature due to the higher density of electronic states at the
Dirac point.
\end{abstract}

\maketitle


\section{Introduction} \label{s.intro}

Two-dimensional graphene in its monolayer and bilayer forms can
exhibit rather different electronic
characteristics.\cite{castroneto09,dassarma10b} In monolayer graphene
(MLG) the valence and conductance bands touch each other at two
inequivalent Dirac points in the Brillouin zone, around which the
bands are linear. Thus the density of states (DOS) also vanishes
linearly around these points, where the Fermi energy of charge-neutral
graphene is located.  In the most common (Bernal) stacking of bilayer
graphene (BLG) the two layers are electronically coupled such that the
two linear bands of the individual layers mix to form four bands, the
lower of which are parabolic around the Dirac points. In
this case the DOS around these points is approximately constant.

This difference gives rise to different charge screening and transport
properties for the two types of graphene.  In particular, the
temperature dependences for the conductivity of diffusive MLG and BLG
around the charge-neutrality point (CNP)
differ.\cite{adam07b,morozov08,adam08,xiao10,adam10,dassarma10}  
In both cases, thermal excitation of quasiparticles from the valence to
the conduction band (i.e.\ thermal creation of electron-like and
hole-like charge carriers) increases the conductivity with
temperature, as is typical for semiconductors and insulators.
However, due to differences in the DOS, in MLG the conductivity at CNP
grows only quadratically with temperature, while in BLG it increases
linearly.\cite{adam08}

In this paper we show how this difference of MLG and BLG is reflected
in the current-voltage ($I(\vds)$) characteristics of diffusive
graphene. For MLG it is known that the $I(\vds)$ characteristics tend
to be linear at low bias voltage $\vds$ and have a tendency to
saturate at higher voltages due to scattering of electrons from
optical phonons.\cite{meric08,barreiro09} Close to CNP the $I(\vds)$
at small $\vds$ can become superlinear as a result of Zener-Klein
tunneling between the valence and conduction bands, especially in
low-mobility samples.\cite{vandecasteele10} By measuring the $I(\vds)$
curves of both MLG and BLG on a SiO$_2$ substrate in a two-terminal
configuration, we show that in BLG the $I(\vds)$ characteristics have
a much stronger tendency for superlinearity at $\vds\lesssim 0.1$ V,
which we associate with an increase of the conductivity due to self
heating (Joule heating). This effect is only weakly dependent on the
level of doping and on the length of the sample.  We confirm this
interpretation with numerical simulations using a semiclassical model
based on Boltzmann theory for a diffusive two-dimensional (2D) system
in quasiequilibrium. The model takes into account electron scattering
from charged impurities, the band-bending effects due to charge doping
by the metallic source and drain
electrodes,\cite{huard08,barraza10,khomyakov10} uniform impurity
doping, as well as nonuniform doping by a gate electrode.  For MLG the
Joule-heating-related nonlinearity is found to be weak, which is
consistent with previous experiments and
calculations.\cite{meric08,barreiro09,vandecasteele10}

The paper is organized as follows. In Sec.\ \ref{s.model} we introduce
the theoretical model, Sec.\ \ref{s.exp} describes the experimental
results and compares them to theory, and in Sec.\ \ref{s.disc} we end
with some discussion of other mechanisms for current nonlinearities.
Details of the model are given in the Appendixes.  In
App.\ \ref{s.green} solution of the electrostatic part of the problem
in terms of a Green function is detailed. Appendix \ref{s.coeffs}
discusses the modeling of the impurity scattering and gives
expressions for the charge density and the transport coefficients for
MLG and BLG.  Finally, in Sec.\ \ref{s.pnj} we discuss analytic
semiclassical results for the temperature dependence of the
conductance of a $p$-$n$ junction, which can form in the graphene
close to the metallic electrodes.


\section{Theoretical model} \label{s.model}

\begin{figure}[!t]
\includegraphics[width=0.7\linewidth,clip=]{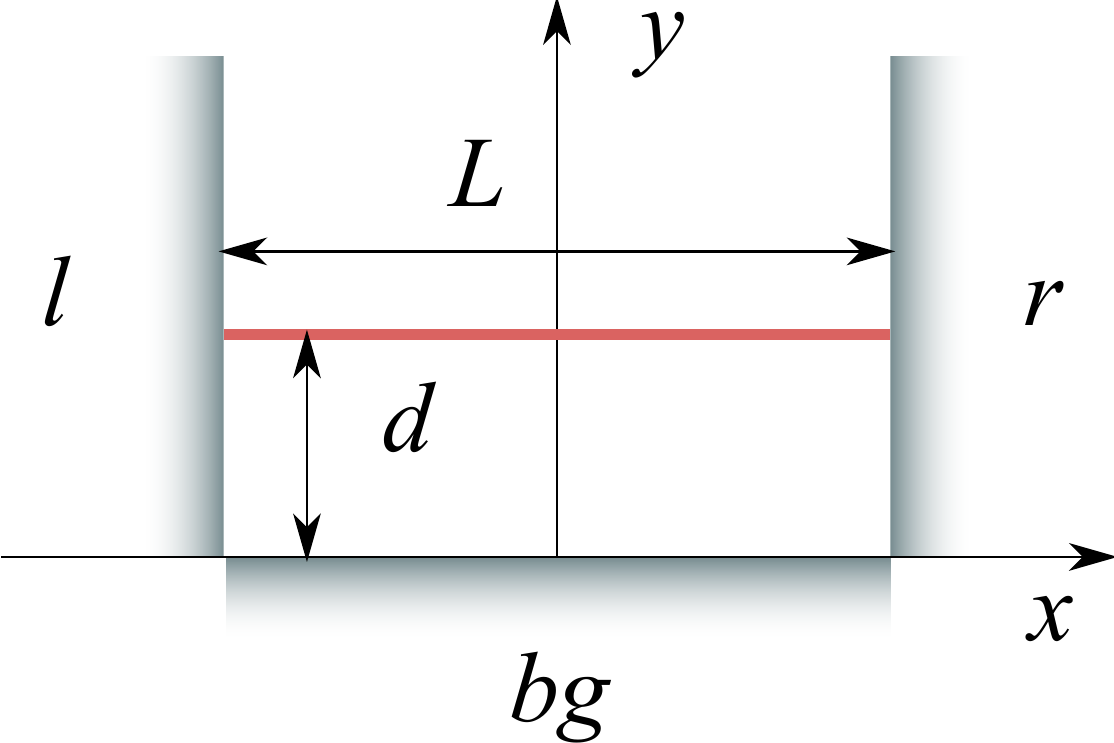}
\caption{The rectangular geometry considered in the model.  $\lel$ and
  $\rel$ depict the left and right (source and drain) electrodes,
  while $\bgel$ is a back gate electrode. All three are held at
  different constant potentials, $\phi_{\lel}$, $\phi_{\rel}$, and
  $\phi_{\bgel}$, respectively.  The 2D channel, of length $\chl$, is
  at a distance $\bgd$ from the back gate.  For simplicity the regions
  below and above the channel are assumed to be occupied by the same
  dielectric medium with permittivity $\varepsilon$.  }
\label{f.geometry}
\end{figure}

The model geometry that we consider (see Fig.\ \ref{f.geometry}) 
is a simplification of typical experimental geometries. It consists of a
two-dimensional (quasi-one-dimensional) channel coupled to two
transport electrodes ($\lel$ and $\rel$) as well as a back gate
electrode ($\bgel$). In this geometry we solve for the electrostatic
potential together with transport equations for the quasiparticles in
the channel.

Our semiclassical transport theory assumes a diffusive 2D electron
system in quasiequilibrium,\cite{viljas10mod} such that the
quasiparticle distribution is described by a local chemical potential
$\mu(x)$ and a local temperature $T(x)$. A third unknown field is the
electrostatic potential $\varphi(x)$ in the channel ($y=d$), which
shifts the energy $E_D(x)$ of the local Dirac point such that
$E_D(x)=-e\varphi(x)$.  The three fields $\varphi(x)$, $\mu(x)$, and
$T(x)$ are solved from the equations
\begin{equation}\begin{split} \label{e.alleqs}
\varphi(x) =& \int_{-\chl/2}^{\chl/2} d\xi G(x,d;\xi,d) e [n(\xi) - n_{dop}]/\varepsilon  \\
&+\sum_{X=\lel,\rel,\bgel} \psi_X(x,d) \phi_X \\
&[\sigma(x)\mu'(x)/e + \gamma(x)T'(x)]' = 0, \\
-&[\alpha(x)\mu'(x)/e + \kappa(x)T'(x)]' =P_J(x).
\end{split}\end{equation}
Here prime denotes the $x$-derivative, $G$ is the Green function of
the Laplace operator, $\psi_X(x,y)$ is the ``characteristic function''
for electrode $X$ (see App.\ \ref{s.green}), and 
$\varepsilon=\varepsilon_r\varepsilon_0$, where 
$\varepsilon_r$ and $\varepsilon_0$ are the relative and vacuum permittivities. 
Further, $n(x)$ is the
two-dimensional charge density in units of the electron charge $-e$
(App.\ \ref{s.coeffs}), and $n_{dop}$ is a phenomenological doping
density that describes the doping effect by impurities, which is
assumed to be constant. (Thus, charge puddles\cite{martin08,deshpande09} are not
described --- see below for discussion). The factors $\sigma(x)$ and
$\kappa(x)$ are the charge and thermal conductivities, respectively,
while $\alpha(x)$ and $\gamma(x)$ are the thermoelectric transport
coefficients, satisfying $\alpha(x)=T(x)\gamma(x)$.  These factors are
inversely proportional to the impurity density $n_{imp}$, which is
taken to be another constant parameter independent of $n_{dop}$
(App.\ \ref{s.coeffs}).  The quantities $n(x)$, $\sigma(x)$,
$\kappa(x)$, $\gamma(x)$, and $\alpha(x)$ all depend on $\varphi(x)$.
Finally, $P_J(x)=j(x)(\mu'(x)/e)$ is the Joule power per area, with
$j(x)=\sigma(x)\mu'(x)/e+\gamma(x) T'(x)$ being the electric current
density, which is conserved, $j(x)\equiv j$.  The total current
through a system of finite width $W\gg L,d$ is $I=jW$.

The boundary conditions at $x=\pm \chl/2$ are chosen to be
\begin{equation}\begin{split} \label{e.bndconds}
\phi_{\lel} &= -V, \quad \phi_{\rel} = 0 \\
\mu(-\chl/2) &= eV + \mu_{eq}, \quad \mu(+\chl/2) = \mu_{eq}  \\
T(-\chl/2) &= T(+\chl/2) = T_0.
\end{split}\end{equation}
which correspond to a voltage bias $V$, assuming negligible contact
resistance, the right-hand electrode to remain grounded, and both
electrodes to remain at the bath temperature $T_0$.  The finite
equilibrium chemical potential $\mu_{eq}\neq 0$ describes the effects
of the work function mismatch and the resulting charge transfer
between the graphene and the electrodes, with $\mu_{eq}>0$
($\mu_{eq}<0$) leading to $n$-type ($p$-type)
doping.\cite{khomyakov10}

The first of Eq.\ (\ref{e.alleqs}) is the Poisson equation written as
an integral equation at $y=d$. The second and third describe current
conservation and heat balance, respectively.  Note that in order to
keep the model simple, we do not include electron-phonon coupling and
thus the Joule power is dissipated only via diffusion.\cite{viljas10mod} This should be
reasonable first approximation at bias voltages $V$ well below optical
phonon energies.\cite{footnote2}

\begin{figure}[!t]
\includegraphics[width=0.9\linewidth,clip=]{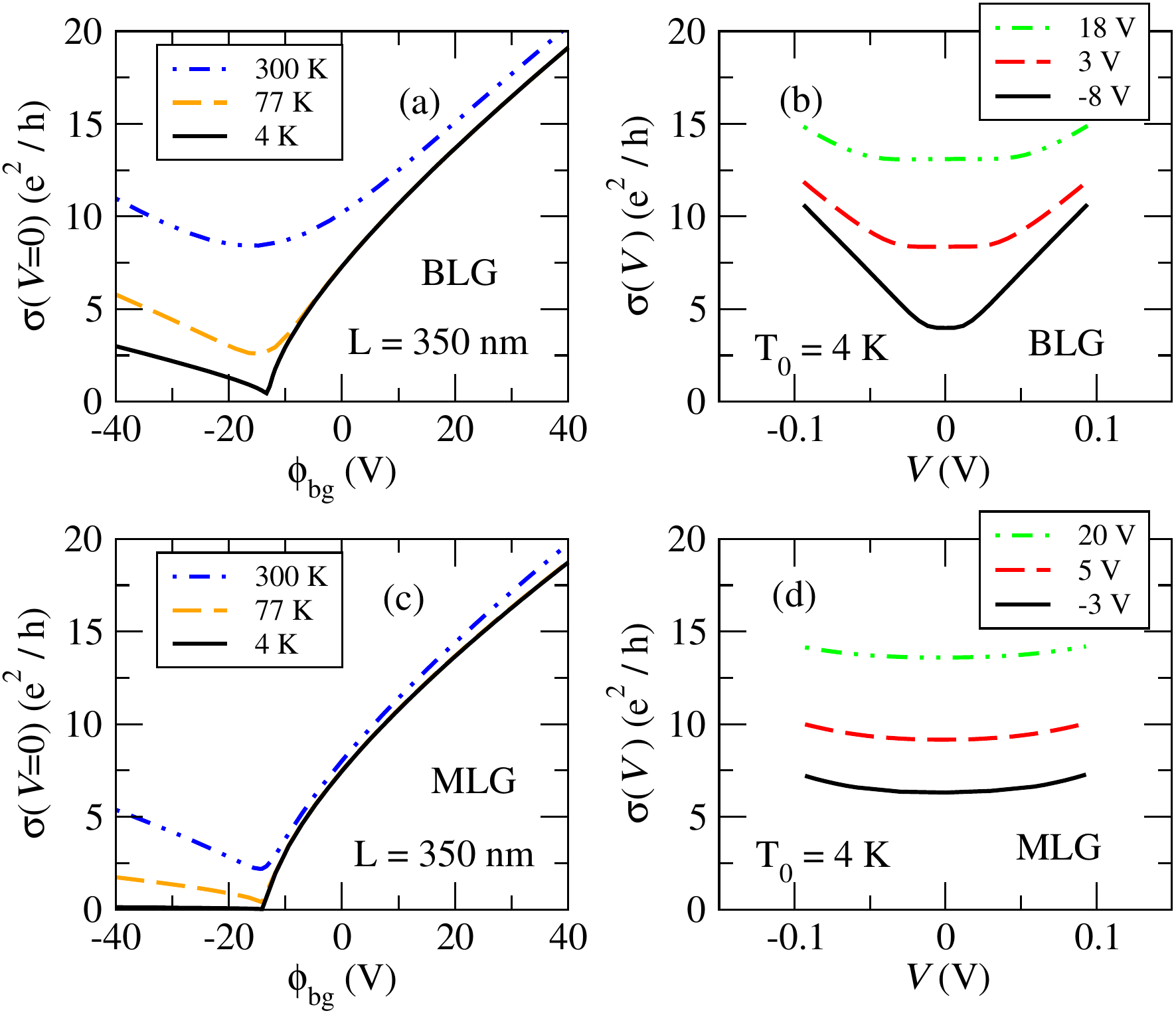}
\caption{ Results for BLG (a,b) and MLG (c,d) with parameter
  $\chl=350$ nm, $\mu_{eq}=0.05$ eV, $n_{dop}=10.0\cdot10^{15}$
  m$^{-2}$, $\varepsilon_r=4.0$, $\bgd=210$ nm.  Additionally for BLG
  $n_{imp}=7.0\cdot10^{16}$ m$^{-2}$, and for MLG
  $n_{imp}=2.5\cdot10^{15}$ m$^{-2}$.  (a,c) Linear-response
  conductivity $\sigma(\vds=0)=(\chl/W)(dI/d\vds)_{\vds=0}$
  at three temperatures and (b,d) 
  the corresponding finite-$V$ differential conductivity 
  $\sigma(\vds)=(\chl/W)(dI/d\vds)$ 
  for $T_0=4$ K
  at indicated gate voltages $\phi_{\bgel}$.
}
\label{f.expcomp_mlgblg}
\end{figure}

\begin{figure}[!t]
\includegraphics[width=0.9\linewidth,clip=]{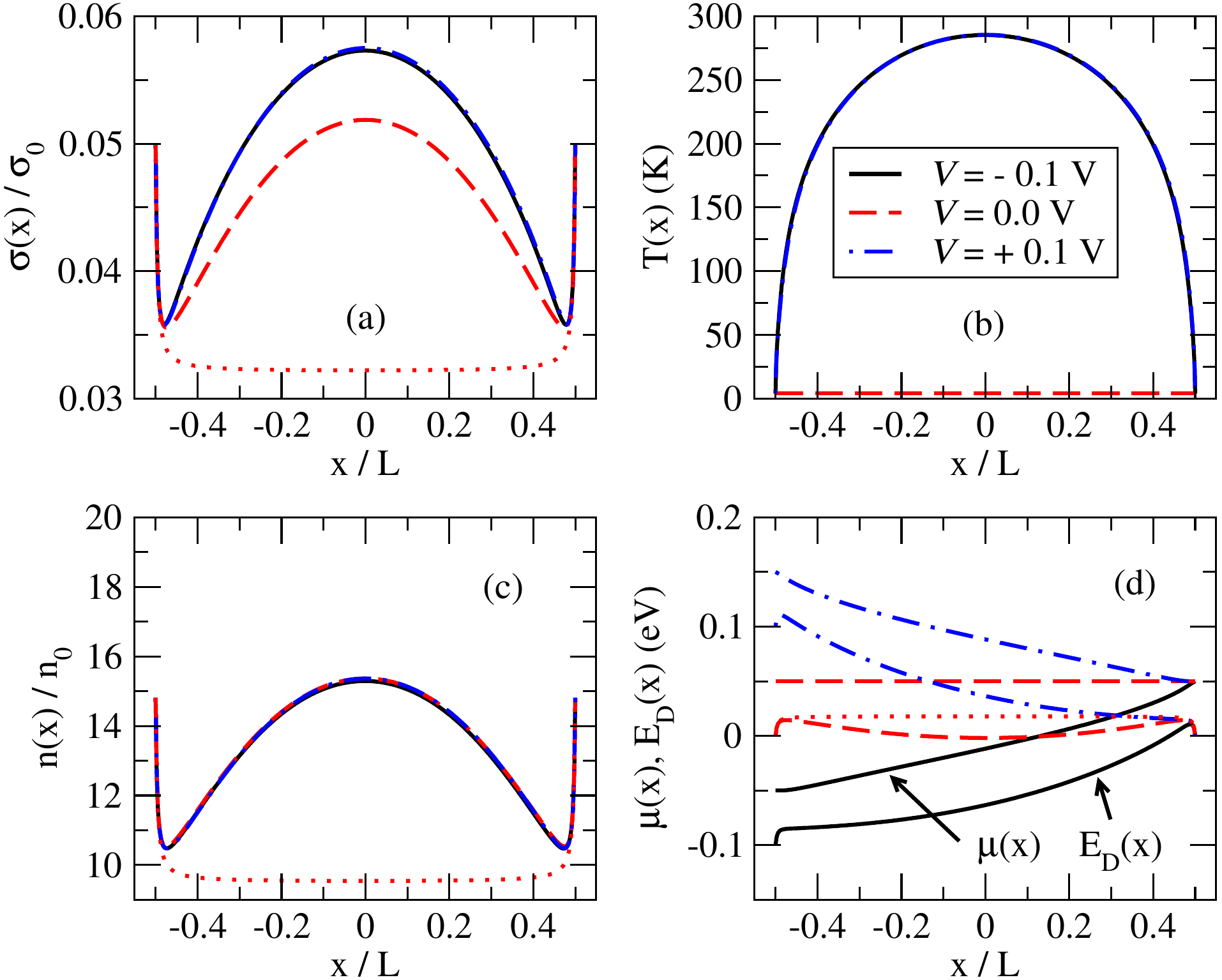}
\caption{Bias dependence of the field profiles for $n$-type BLG at gate voltage
  $\phi_{\bgel}=8$ V: (a) conductivity $\sigma$, (b) temperature $T$
  (c) charge density $n$ (d) chemical potential $\mu$ and the local
  Dirac point $E_D$ (higher and lower curve, respectively).  Solid, dashed,
  and dash-dotted curves are for $\vds=-0.1$, $0.0$, and $0.1$ V,
  respectively.  Dotted lines show additionally $\sigma$, $n$, and
  $E_D$ in equilibrium at $\phi_{\bgel}=0$; here $n(x=0)\approx n_{dop}$.  
  The parameters are as in
  Fig.\ \ref{f.expcomp_mlgblg}, and we have defined the units
  $\sigma_0=C_{BLG}E_0$ and $n_0=(\varepsilon/d) E_0/e^2$, where
  $E_0=$1 eV (for $C_{BLG}$, see App.\ \ref{s.coeffs}).  }
\label{f.expcomp_blg_profs}
\end{figure}

In Fig.\ \ref{f.expcomp_mlgblg} we show typical results obtained from
the model for BLG and MLG with parameters obtained from the fit to
(the gate dependence of) our BLG experiments in Sec.\ \ref{s.exp}.  It
is to be noted that the aspect ratio $\chl/\bgd=1.7$ is far too small
for a simplified parallel-plate capacitor model to work properly in
the geometry of Fig.\ \ref{f.geometry}, and a full numerical solution
of Eqs.\ (\ref{e.alleqs}) is needed.  The fields obtained as their
solutions are very nonuniform, as shown for BLG in
Fig.\ \ref{f.expcomp_blg_profs}.

The upper panels of Fig.\ \ref{f.expcomp_mlgblg} are for the case of
BLG.  Fig.\ \ref{f.expcomp_mlgblg}(a) shows the gate dependence of the
linear-response conductivity
$\sigma(\vds=0)=(\chl^{-1}\int_{-\chl/2}^{\chl/2}\rho(x')dx')^{-1}$,
where $\rho(x)=1/\sigma(x)$ is the resistivity.  Due to finite
positive doping density ($n_{dop}>0$) the point
of minimal conductivity (the apparent ``CNP'') 
is shifted to $\phi_{\bgel}=\phi_{\bgel,min}\approx -13$ V
and as result of the lead-doping effect ($\mu_{eq}>0$), the gate
dependence is asymmetrical around the CNP.\cite{huard08} Since the
lead-doping is of $n$ type, at $\phi_{\bgel}>\phi_{\bgel,min}$ the BLG
is of $n$ type everywhere. However, for
$\phi_{\bgel}<\phi_{\bgel,min}$ two $p$-$n$ junctions
appear,\cite{huard08} and the BLG becomes of $n$-$p$-$n$ type. While
for $\phi_{\bgel}>\phi_{\bgel,min}$ the dependence on temperature is
relatively weak, for $\phi_{\bgel}\approx \phi_{\bgel,min}$ it is
roughly linear: $\sigma(\vds=0)\propto T$ (Ref.\ \onlinecite{adam08}).
For $\phi_{\bgel}\ll \phi_{\bgel,min}$, where the $p$-$n$ junctions
dominate the resistance, it can be shown that approximately
$\sigma(\vds=0)\propto -1/\ln(T)$ as $T\rightarrow 0$
(App.\ \ref{s.pnj}).  However, the theory is valid only when all parts
of graphene remain far from the local CNP. This is because charge
puddles, quantum-mechanical effects (see App.\ \ref{s.pnj}), and a
possible gap in the BLG spectrum are not taken into account.  Thus, we
mainly concentrate on the gate voltages $\phi_{\bgel}>\phi_{\bgel,min}$.

Fig.\ \ref{f.expcomp_mlgblg}(b) shows the differential conductivity
$\sigma(\vds)=(\chl/W)(dI/d\vds)=\chl(dj/d\vds)$ as a function of
$\vds$ for a few gate voltages $\phi_{\bgel} > \phi_{\bgel,min}$ at
low bath temperature.  Note that the $\sigma(\vds)$ curves are nearly
symmetric [$\sigma(-V)\approx\sigma(V)$], although there are small
deviations, which are due to the asymmetrical choice of the boundary
conditions and the presence of the gate electrode.  The increase of
$\sigma(\vds)$ at finite $\vds$ signifies a superlinear contribution
to the $I(\vds)$ curve: $I(\vds)\approx G_1\vds+G_2\vds^2$ ($\vds>0$)
with $G_2>0$.  This superlinearity is strongest close to the CNP,
where the conductivity of BLG is most sensitive to temperature (see
Eq.\ (\ref{e.blgtrcoeffs})).  In fact, the increase of the
conductivity with voltage [see Fig.\ \ref{f.expcomp_blg_profs}(a)] is
entirely due to heating, which leads to maximal temperatures of
$T(x)\sim 300$ K at $\vds=0.1$ V [Fig.\ \ref{f.expcomp_blg_profs}(b)].
The charge density $n(x)$ for example, remains almost independent of
$V$ [Fig.\ \ref{f.expcomp_blg_profs}(c)].  Indeed, the density is of
the form $n(x)=\hat n(\mu(x)-E_D(x))$ (App.\ \ref{s.coeffs}), and
$\mu(x)-E_{D}(x)$ remains close to its value at $V=0$ everywhere
[Fig.\ \ref{f.expcomp_blg_profs}(d)].  Thus the bias voltage ``gates''
the graphene very little.  The dotted lines in
Fig.\ \ref{f.expcomp_blg_profs} additionally show the results at
equilibrium, with $\phi_{\bgel}=0$. The fast transients close to the
electrodes are due to the doping by the leads.  This doping is not
restricted only to the region on the order of a screening length $\sim
1$ nm (Sec.\ \ref{s.coeffs}) from the leads, but is actually
long-ranged.\cite{shikin01,khomyakov10}

The lower panels Fig.\ \ref{f.expcomp_mlgblg}(c,d) show equivalent
results for MLG, where the impurity density $n_{imp}$ has been chosen
so that the conductivities are of a similar magnitude as for BLG. The
temperature-dependence of $\sigma(\vds=0)$ for
$\phi_{\bgel}\gtrsim\phi_{\bgel,min}$ is now clearly even weaker.
For $\phi_{\bgel}\approx\phi_{\bgel,min}$ it is quadratic,
$\sigma(\vds=0)\propto T^2$ (Ref.\ \onlinecite{adam08}), and for 
$\phi_{\bgel}\ll\phi_{\bgel,min}$ linear,
$\sigma(\vds=0)\propto T$ (App.\ \ref{s.pnj}).
Correspondingly, the increase of the
$\sigma(\vds)$ at $\phi_{\bgel}>\phi_{\bgel,min}$ is much weaker. 
This is consistent with the fact that
the $I(\vds)$ curves measured for MLG are typically linear or even
sublinear, except close to CNP in low-mobility samples, where Zener-Klein
tunneling is of importance.\cite{vandecasteele10} In the case of MLG
the ``gating'' effect of the bias voltage is somewhat larger due to
the longer screening length, but remains also weak.

Here we have concentrated on short samples, with $\chl/\bgd\sim 1$. In
the considered model geometry the gate dopes the graphene quite weakly
at distances on the order of $d$ from the ends. Additionally, the
center of the graphene heats more than the ends.  Therefore at
$\phi_{\bgel}>\phi_{\bgel,min}$ the ends tend to dominate the
resistance [Fig.\ \ref{f.expcomp_blg_profs}(a)].  When $\chl\gg \bgd$,
the parallel-plate limit is approached, where $n(x)$ and $\sigma(x)$
become uniform, with $\mu(x)$ and $E_D(x)$ roughly linear.  One may
then simplify the equations Eqs.\ (\ref{e.alleqs}) by taking
$\alpha=\gamma=0$ and using the Wiedemann-Franz law
$\kappa=\mathcal{L}T\sigma$, where $\mathcal{L}=(\pi^2/3)(k_B/e)^2$,
and assuming a constant $\sigma$. The temperature profile is thus
approximated with $T(x)=\sqrt{T_0^2+[1/4-(x/\chl)^2]V^2/\mathcal{L}}$,
which scales simply with $V$.  The heating effect on the conductivity
$\sigma(V)$ therefore depends relatively weakly on the length of the
sample.

We do not pursue further simplifications or extensions of the model
here, but it should be noted that the thermoelectric coefficients
$\alpha$ and $\gamma$ are not of great importance for the current
nonlinearity.  However, under some conditions the strong temperature
gradient at the ends can also cause the conductivity to decrease at
small bias voltage.  A very weak sign of this is seen in the flat
region of the $\phi_{\bgel}=18$ V curve in
Fig.\ \ref{f.expcomp_mlgblg}(b).  We also note that our tests with
some simple models for charge puddles can reduce the width of this
flat region, making nonlinearity stronger also at high gate voltages.


\section{Experiments} \label{s.exp}

\begin{figure}[!t]
\includegraphics[width=0.85\linewidth,clip=]{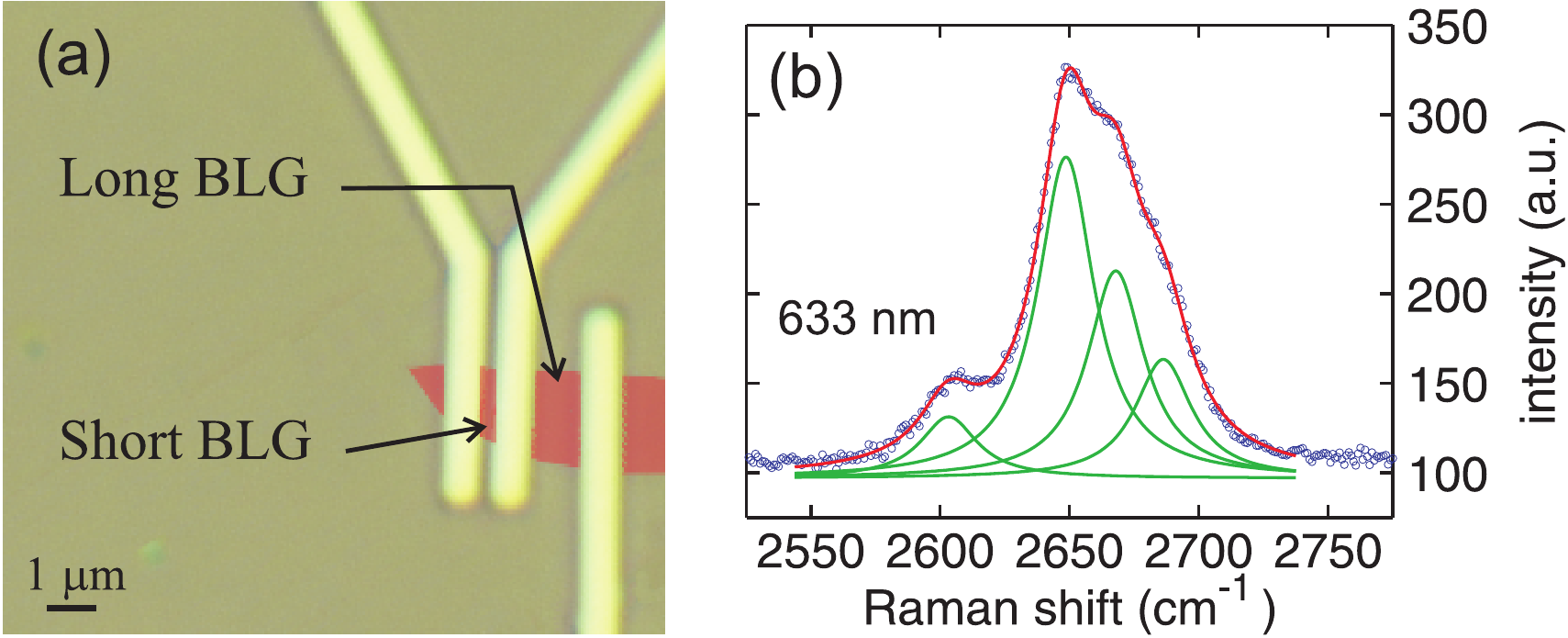}
\caption{(a) Optical picture of two bilayer graphene samples of
  lengths $\chl=350$~nm and $\chl=950$~nm.  The BLG flake has been colored
  in red to enhance its visibility.  (b) Raman spectrum of the
  corresponding bilayer graphene sheet.\cite{ferrari06} 
}
\label{f.raman_geom}
\end{figure}

We have measured seven bilayer graphene samples in a two-lead
configuration and found qualitatively the same transport properties
for each of them.  Here we focus on the results obtained on two
samples from the same BLG sheet having lengths $\chl=350$ nm and
$\chl=950$ nm, and widths $W=900$ nm and $1550$ nm, respectively
(Fig.\ \ref{f.raman_geom}).  The samples were contacted using Ti/Al/Ti
sandwich structures with thicknesses 10 nm / 70 nm / 5 nm (10 nm of Ti
is the contact layer). Three 0.6 $\mu$m wide contacts were patterned
using e-beam lithography. The strongly doped Si substrate was used as
a back-gate, separated by 270 nm of SiO$_2$ from the sample.

\begin{figure}[!t]
\includegraphics[width=0.75\linewidth,clip=]{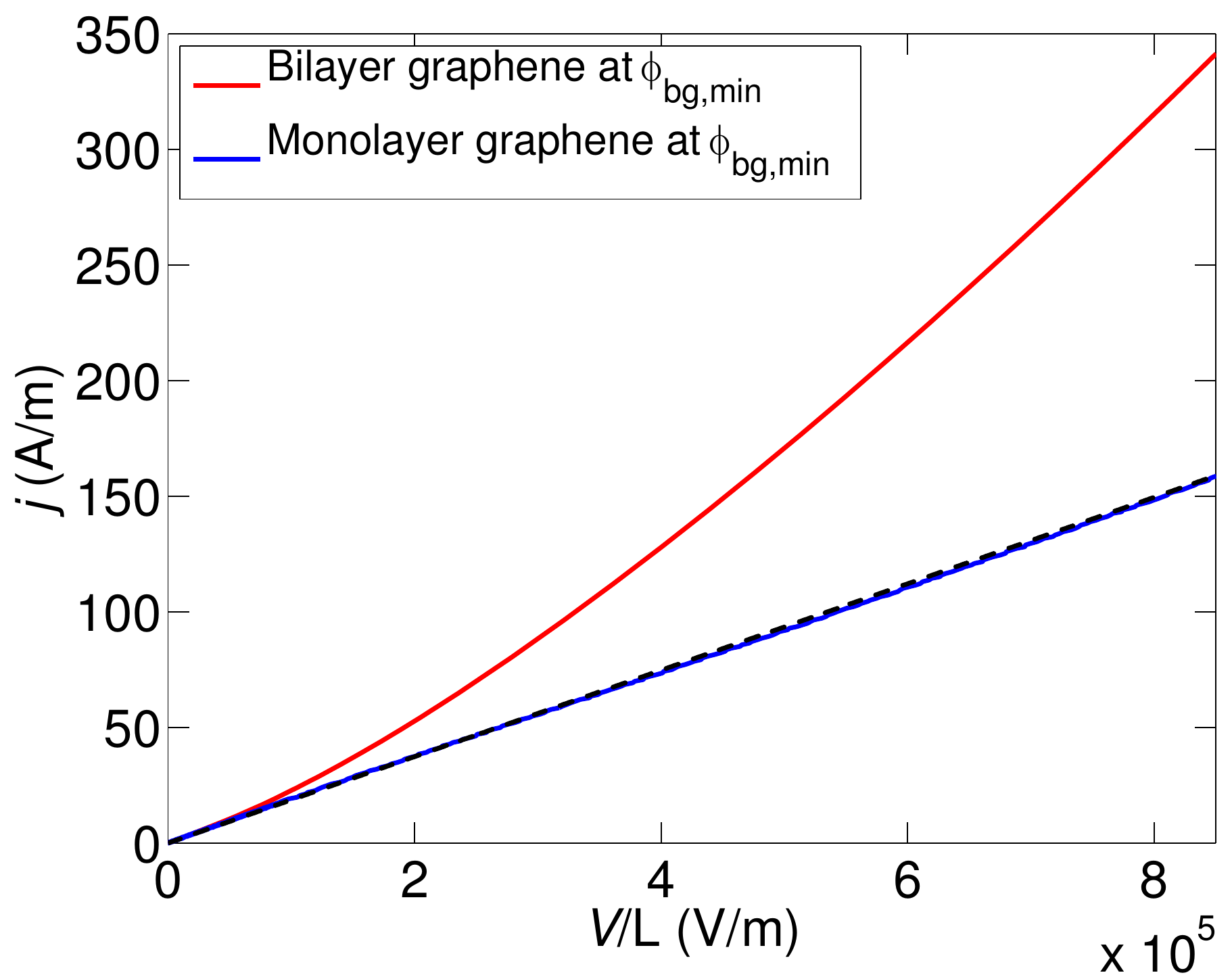}
\caption{ Typical $I(\vds)$ curves for MLG and BLG at the CNP at $4.2$ K. 
  The dashed line has a slope equal to the conductivity $4.5e^2/h$ of both
  samples at $V=0$. For MLG the $I(\vds)$ curve is linear.  In BLG it
  is superlinear, which we associate with the Joule heating of the
  sample.  }
\label{f.iv_blg_mlg}
\end{figure}

\begin{figure}[!t]
\includegraphics[width=0.95\linewidth,clip=]{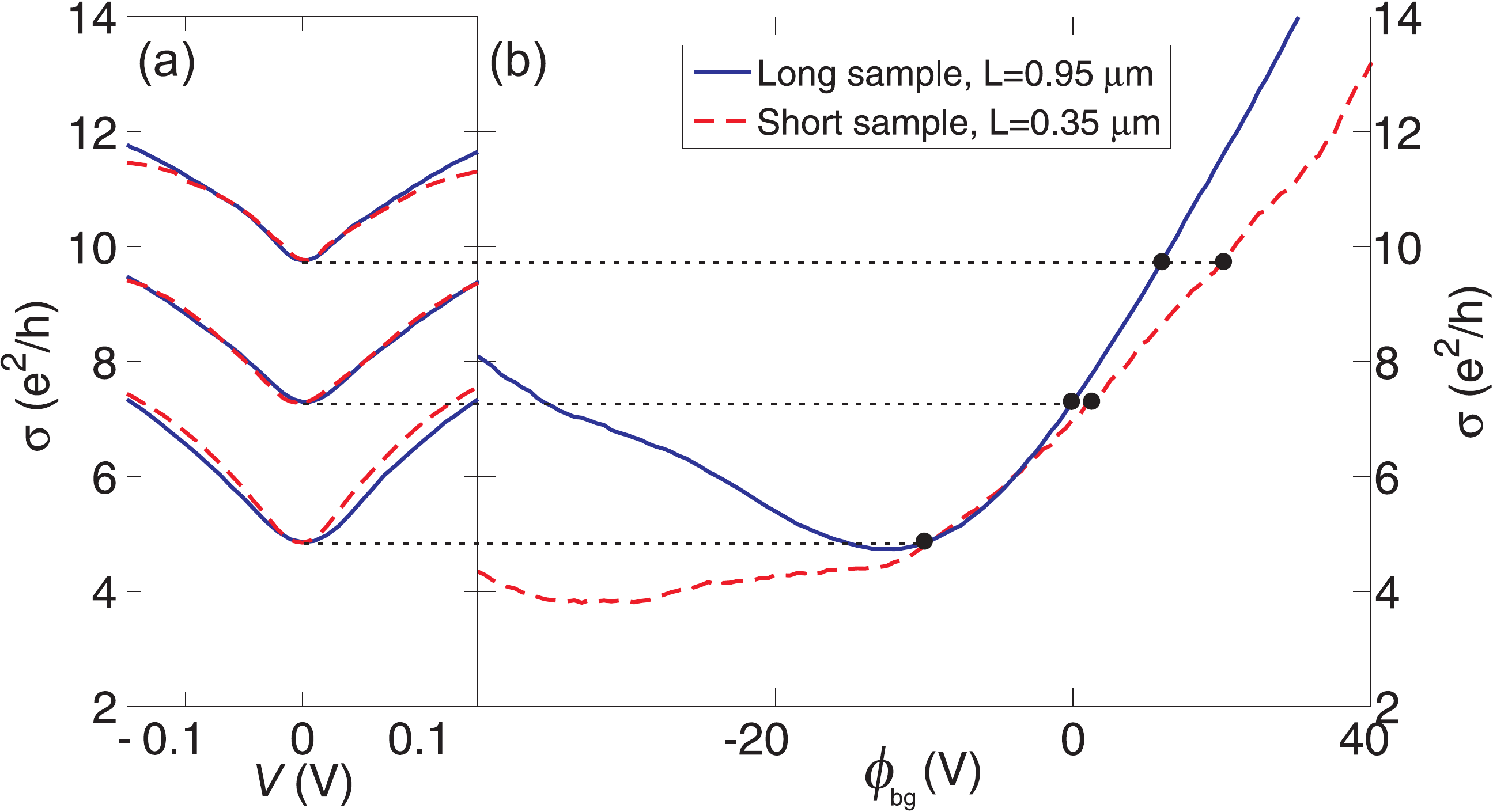}
\caption{ 
  The left-hand panel (a) shows the experimentally measured
  differential conductivity $\sigma(\vds)$ at finite $\vds$ for both
  the short (dashed line) and the long (solid line) BLG sample at
  three pairs of gate voltages that are chosen so that the minima at
  $\vds=0$ V for both samples coincide.  The right-hand panel (b)
  shows the corresponding zero-bias conductivity $\sigma(\vds=0)$
  vs.\ the gate voltage $\phi_{\bgel}$ for both samples. The
  temperature is $T_0=4.2$ K.  }
\label{f.long_short}
\end{figure}

The $I(\vds)$ curve of our 0.95 x 1.55 $\mu$m$^2$ sample at $4.2$ K is
illustrated in Fig.\ \ref{f.iv_blg_mlg} together with the $I(\vds)$ in
a typical MLG sample.  While the MLG result is
linear,\cite{meric08,barreiro09} the BLG curve exhibits clearly
superlinear behavior at small drain-source voltage $V$; the
nonlinearity in BLG depends relatively weakly on the gate voltage
$\phi_{\bgel}$ but is largest near the CNP. The nonlinear behavior can
be observed more clearly in Fig.\ \ref{f.long_short}(a) which displays
the differential conductivity $\sigma(\vds)=(\chl/W)(dI/d\vds)$ of the
long and short samples at a few values of $\phi_{\bgel}$.  It is seen
that also the length dependence of the nonlinearity at bias voltages
below $V\approx 0.1$ V is weak.  This supports its interpretation as a
heating effect: as mentioned above, the temperature should scale with
$V$ and not, for example, the electric field $V/L$.  Note,
furthermore, that the $\sigma(V)$ curves are very symmetrical.

Figure \ref{f.long_short}(b) shows the full gate voltage dependence of
the zero-bias conductivity of the short and long samples. In both
samples, the minimal conductivity is located in the negative gate
voltage region around -10 V.  The minimum zero-bias conductivities are
roughly $3.8$ and $4.7$$e^2/h$ for the short and long sample,
respectively.  These are close to the value $\sim 4e^2/h$ typically
found for both MLG and BLG.\cite{geim07,morozov08} An asymmetry
between the $n$-doped and $p$-doped regions is clearly visible and is
more pronounced for the short sample where the conductivity is almost
constant in the $p$-region.  We interpret this electron-hole asymmetry
as a sign of the leads $n$ doping the
graphene,\cite{huard08,khomyakov10} so that there are $p$-$n$
junctions present at larger negative gate voltages.  This is
consistent with expectations for Ti/Al
electrodes.\cite{huard08,khomyakov10} In a parallel-plate
approximation the charge density is $n\approx
(C_{\bgel}/e)(\phi_{\bgel}-\phi_{\bgel,min})$, where
$C_{\bgel}=1.3\cdot 10^{-4}$ F/m$^2$. Using this we estimate from the
slope of $\sigma(V=0)$ vs.\ $n>0$ for the long sample the mobility
$\mu_{m}=\sigma/(en)$ to be at least $1500$ cm$^2$V$^{-1}$s$^{-1}$.
Using this the mean free path is estimated to be $l_{mfp}=\sqrt{\pi
  n}\hbar\mu_{m}/e\lesssim 30 $ nm, and thus the samples are
diffusive.

\begin{figure}[!t]
\includegraphics[width=0.9\linewidth,clip=]{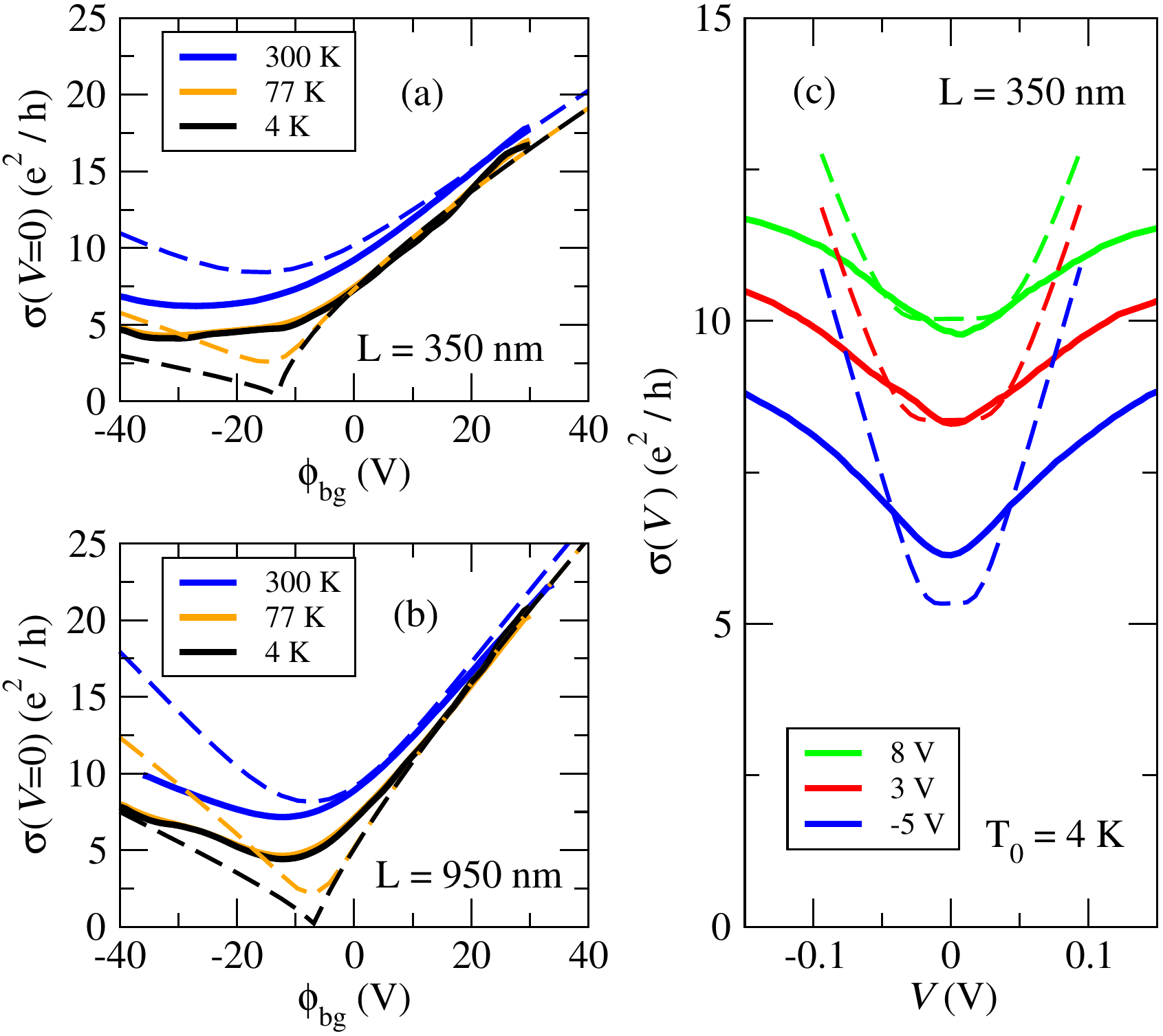}
\caption{
  Experimental results (solid lines) and a theoretical fit
  (dashed lines) for a short BLG sample ($\chl=350$ nm) and a long BLG
  sample ($\chl=950$ nm).  Left-hand panels (a,b) show the linear
  response conductivity at $T=300$ K (top curve), $77$ K, and $4$ K (bottom curve), 
  and the right-hand panel (c) shows the differential conductivity 
  for $\chl=350$ nm at $T_0=4$ K 
  and at gate voltages $8$ V (top curve), $3$ V, and $-5$ V (bottom curve).
  The parameters used for the
  theoretical fit are: $\mu_{eq}=0.05$ eV, $n_{imp}=7.0\cdot10^{16}$
  m$^{-2}$, $\varepsilon_r=4.0$, $\bgd=210$ nm.  For the short sample
  $\chl=350$ nm and $n_{dop}=10.0\cdot10^{15}$ m$^{-2}$, while for the
  long sample $\chl=950$ nm, and $n_{dop}=7.0\cdot10^{15}$ m$^{-2}$.
}
\label{f.expcomp_blg_fit}
\end{figure}

We compare the raw experimental data to the theoretical results in
Fig.\ \ref{f.expcomp_blg_fit}, where solid and dashed lines are for
experimental and theoretical data, respectively.  Figures
\ref{f.expcomp_blg_fit}(a) and (b) show the $\sigma(\vds=0)$
vs.\ $\phi_{\bgel}$ dependence for the short and the long sample,
respectively.  Also experimental data measured at 77 K and 300 K are
shown.  It is seen that the zero-bias conductivity increases slightly
from 4 K to 77 K, and more significantly from 77 K to 300 K. The
conductivity change is strongest near the CNP and also in the
$p$-doped region.  Since the theory is expected to work only in the
absence of $p$-$n$ junctions, the fitting is done only to the
$\sigma(\vds=0)$ vs.\ $\phi_{\bgel}$ data
[Fig.\ \ref{f.expcomp_blg_fit}(a,b)] at
$\phi_{\bgel}>\phi_{\bgel,min}$, with $\mu_{eq}>0$. The same
parameters are then used for calculating $\sigma(\vds)$. The
$\sigma(\vds)$ data for the short sample are shown in
Fig.\ \ref{f.expcomp_blg_fit}(c). 

The parameters used for the fit are given in the caption of
Fig.\ \ref{f.expcomp_blg_fit}.  The work function mismatch
$\mu_{eq}=0.05$ eV is of the correct sign and order expected for Ti/Al
electrodes.\cite{khomyakov10} A gate distance $\bgd=210$ nm is used,
which is smaller than the experimental $270$ nm. The smaller $d$ used
for the comparison is reasonable, since in the simplified geometry
assumed by the theoretical model the gate potential is more strongly
screened by the transport electrodes: even for the long sample the
parallel plate limit is not fully reached.\cite{footnote3} It is also
notable that the average doping density $n_{dop}=7$--$10\cdot10^{15}$
$1/$m$^2$ is much smaller than the impurity density
$n_{imp}=7.0\cdot10^{16}$ $1/$m$^2$, although the simplest theories
for impurity doping predict these two to be equal.\cite{adam08} Even
though we model, for simplicity, the impurities to be all similar and
of the screened-Coulomb type (App.\ \ref{s.coeffs}), in reality there
may be several different types of disorder\cite{xiao10} present in our
samples, which complicates the situation.  The orders of magnitude
$n_{dop},n_{imp}\sim 10^{15}$--$10^{16}$ $1/$m$^2$, are still in the
range that can be expected from other
experiments.\cite{morozov08,xiao10}

The overall agreement of the theory with the experiment is good, apart
from the large deviations for $\phi_{\bgel}\lesssim\phi_{\bgel,min}$,
where some parts of the system are close to the CNP.  At these gate
voltages the low-temperature theoretical results for $\sigma(\vds=0)$
tend to fall well below the experimental ones, whereas at 300 K the
opposite is true.  In addition to this the clearest discrepancy is
that the self-heating predicted by the theory is too strong, as shown
by the overly steep slope of $\sigma(\vds)$ in
Fig.\ \ref{f.expcomp_blg_fit}(c).  Also, the slight decrease of the
experimental slope with voltage is not captured.  A proper explanation
of these effects would require considering interactions of the
electrons with phonons, particularly the remote interface phonons of
the SiO$_2$ substrate.\cite{viljas10mod,footnote2} Another clear
deviation is that the theoretical $\sigma(\vds)$ curves for high
$\phi_{\bgel}$ tend to be flatter close to $\vds=0$ than in the
experiments.  (The theoretical results for the long sample are
otherwise similar as in Fig.\ \ref{f.expcomp_blg_fit}(c), but even
slightly more ``flat''.)  As suggested in Sec.\ \ref{s.model}, the
correct shape could presumably be reproduced by considering charge
puddles, i.e., a non-uniform impurity density and doping. To keep the
model simple and the number of fitting parameters small, we have
neglected puddles here.

It should be noted the gross features of the results can be understood
simply based on the temperature-dependence of the local
conductivity,\cite{adam08} which may be worked out analytically
[Eq.\ (\ref{e.blgtrcoeffs}) below] and the fact that the average
temperature increases roughly linearly with the bias voltage.
However, the precise shape of the nonlinearity is dependent on the
various sources of nonuniformity.


\section{Discussion} \label{s.disc}

The heating effects described above are not the only possible sources
of nonlinearity.  As already mentioned, at high enough bias voltage
electron scattering from phonons tends to reduce the conductivity,
which is expected to make the current-voltage curves at high bias
sublinear.  This has been seen in MLG as a tendency for the current to
almost saturate.\cite{meric08,barreiro09} We also see similar effects
in our experiments\cite{fay09v1} with BLG at voltages $\vds\gtrsim
0.1$ V.

The possibility of nonlinear $I(V)$s in graphene at low bias have also
been discussed based on simple arguments involving the
energy-dependence of the number of open transport channels in the
Landauer-B\"uttiker description of the linear-response
conductivity.\cite{blanter07,sonin08,fay09v1} Such calculations are
problematic for the prediction of current-voltage characteristics
beyond linear response, however, since they do not consider the role
of the actual voltage profiles.\cite{blanter07} Ideally, the
electrostatic potential drop should be calculated self consistently,
as we have done above.

Other mechanisms for nonlinear (mostly superlinear) current-voltage
responses in graphene have very recently been discussed by many
authors.\cite{vandecasteele10,dora10,rosenstein10,wangzheng10,zhou10,joung10,bistritzer10}
In particular, Zener-Klein tunneling in MLG has been shown to give
rise to superlinearities close to
CNP.\cite{vandecasteele10,dora10,rosenstein10} It seems unlikely,
however, that a similar mechanism would be of importance in our
experiments, since the nonlinearity is weakly gate-dependent.  Other
possibilities for nonlinearities include the presence of tunnel
junctions\cite{wangzheng10} or contact phenomena,\cite{zhou10} but we
can disregard them as an explanation of our measurements due to the
weak length-dependence of the nonlinear conductivity $\sigma(V)$.
Furthermore, nonlinear current-voltage curves in graphene oxides have
been explained with space-charge limited currents,\cite{joung10} but
such effects are likely to be negligible in metallic graphene, as also
supported by the absence of any bias-doping effects in our
simulations.  Nonlinearities are predicted also for vertical transport
in misaligned BLG or few-layer systems.\cite{bistritzer10} However, of
all these possibilities the self-heating scenario presented above
seems to be the most likely explanation of our experimental results.

To summarize, we have shown how Joule heating can contribute to the
shape of the observed current-voltage characteristics of bilayer
graphene in the diffusive limit. The heating is signified by a strong
superlinear contribution in $I(\vds)$ and thus a low-bias differential
conductivity $\sigma(\vds)$ increasing with $\vds$.  Our experimental
results and our numerical calculations are in good overall agreement
for bias voltages $V\lesssim 0.1$ V.  The heating effect is much
stronger in bilayer graphene than in monolayer, as can be expected
from the differences in their electronic structures.

\begin{acknowledgments}
We thank 
R. Danneau,
M. Y. Tomi, J. Wengler, F. Wu, and R. H\"anninen
for fruitful discussions. 
This work was supported by the Academy of Finland,
the European Science Foundation (ESF) under
the EUROCORES Programme EuroGRAPHENE,
and the CIMO exchange grant KM-07-4656 (MW).
\end{acknowledgments}


\appendix

\section{Green function and characteristic functions} \label{s.green}

The Green function $G(x,y;x',y')$ of the Laplace operator
satisfies $\nabla^2G(x,y;x',y') = \delta(x-x')\delta(y-y')$,
$G(x',y';x,y)=G(x,y;x',y')$ and zero boundary conditions
on the electrodes. 
For the particular geometry 
under consideration (Fig.\ \ref{f.geometry}) the Green function may be found analytically
using conformal mapping techniques and it is 
\begin{widetext}
\begin{equation}\begin{split} \label{e.greenf}
G(x,y;x',y') = \frac{1}{2\pi}\ln\sqrt{
\frac{
(\sin\td{x}\cosh\td{y} - \sin\td{x}'\cosh\td{y}')^2
+(\cos\td{x}\sinh\td{y} - \cos\td{x}'\sinh\td{y}')^2
}
{
(\sin\td{x}\cosh\td{y} - \sin\td{x}'\cosh\td{y}')^2
+(\cos\td{x}\sinh\td{y} + \cos\td{x}'\sinh\td{y}')^2,
}
}
\end{split}\end{equation}
\end{widetext}
where $\td{x}=\pi x/\chl$, $\tilde{y} = \pi y /\chl$ and so on.

The characteristic function $\psi_X$ for electrode $X$
is defined such that it
satisfies the Laplace equation $\nabla^2\psi_X=0$ with 
the boundary condition that $\psi_X=1$ on electrode $X$
and $\psi_X=0$ on the other electrodes. 
Once the Green function is known, the characteristic functions 
may be easily represented in terms of it:
\begin{equation}\begin{split}
\psi_{\lel}(x,y) &= -\int_0^\infty dy' \frac{\partial}{\partial x'}G(x,y;x',y')\big|_{x'=-\chl/2}  \\
\psi_{\rel}(x,y) &= \int_0^\infty dy' \frac{\partial}{\partial x'}G(x,y;x',y')\big|_{x'=\chl/2}  \\
\psi_{\bgel}(x,y) &= -\int_{-\chl/2}^{\chl/2} dx' \frac{\partial}{\partial y'}G(x,y;x',y')\big|_{y'=0}.
\end{split}\end{equation}


\section{Charge concentration and transport coefficients for graphene} \label{s.coeffs}

In our semiclassical model the charge density 
is assumed to be of the form
\begin{equation}\begin{split} \label{e.ndef}
n(x) 
=& \int_{-\infty}^{\infty} dE \mathcal{D}_\varphi(E,x) 
\big[f(E - \mu(x),T(x)) \\
&- \theta(E_D(x)-E)\big], \\
\end{split}\end{equation}
where $f(E,T)=1/(e^{E/k_BT}+1)$ is the Fermi function,
$E_D(x)=-e\varphi(x)$,
and we define
$\mathcal{D}_\varphi(E,x)=\mathcal{D}(E+e\varphi(x))$,
where $\mathcal{D}(E)$ is the density of states (DOS).
Using Eq.\ (\ref{e.ndef}) self consistently in the Poisson equation 
is essentially the Thomas-Fermi approximation.\cite{khomyakov10}

Assuming a diffusive system with only elastic impurity scattering
the transport coefficients are given by
\begin{equation}\begin{split} \label{e.coeffdefs}
\sigma(x) = & \int dE \sigma_\varphi(E,x) F(E-\mu(x),T(x)) \\
\kappa(x) = & \frac{k_B}{e^2}\int dE \sigma_\varphi(E,x)
\frac{(E-\mu(x))^2}{k_BT(x)} F(E-\mu(x),T(x)) \\
\gamma(x)=&\frac{k_B}{e}\int dE \sigma_\varphi(E,x)
\frac{E-\mu(x)}{k_BT(x)} F(E-\mu(x),T(x)), \\
\end{split}\end{equation}
with $\alpha(x)=T(x)\gamma(x)$.
Here $F(E,T)=-\partial f(E,T)/\partial E$
is the thermal broadening function, and we define
$\sigma_\varphi(E,x)=\sigma(E+e\varphi(x))$,
where
$\sigma(E)=e^2D(E)\mathcal{D}(E)$ is the energy-dependent Boltzmann 
conductivity.
The quantity $D(E)=v^2(E)\tau(E)/2$ the diffusion constant, where
$v(E)$ is the group velocity and $\tau(E)$ the transport
relaxation time.

It is easy to see by change of the integration variable that
the quantities in Eqs.\ (\ref{e.ndef}) and (\ref{e.coeffdefs}) 
only depend on the difference $\mu(x)-E_D(x)$. Thus 
below we define the ``hatted'' quantities with 
$n(x)=\hat{n}(\mu(x)-E_D(x),T(x))$,
$\sigma(x)=\hat{\sigma}(\mu(x)-E_D(x),T(x))$, and similarly 
for the other transport coefficients. 

As a specific model for the impurity scattering 
we consider only screened Coulomb impurities, which 
lead to a linear dependence of the conductivity on charge density\cite{adam08}
for both BLG and MLG, as observed in most experiments.\cite{morozov08}
(For BLG also short-range scattering may be of importance.\cite{xiao10})
For simplicity we assume all of the bare impurities to carry 
a charge $\pm e$ and to be at zero distance from the 
graphene, and perform an average with respect
to their positions.\cite{adam08} 
For BLG and MLG some further approximations
are made, as explained below.

\subsection{Bilayer graphene} \label{s.blgcoeffs}

For BLG
we assume a purely parabolic and gapless dispersion
$E=\pm(\hbar v_0 k)^2/\gamma_1$, where $v_0=10^6$ m/s and $\gamma_1=0.4$ eV.
This yields a group velocity $v(E)=2v_0\sqrt{|E|/\gamma_1}$ and
a constant DOS $\mathcal{D}(E)=\frac{1}{\pi}\frac{\gamma_1}{(\hbar v_0)^2}$.
Then
\begin{equation}\begin{split}
\hat{n}(\mu,T) = \frac{1}{\pi}\frac{\gamma_1}{(\hbar v_0)^2}\mu.
\end{split}\end{equation}
The DOS leads to an inverse Thomas-Fermi screening length
$q_{TF,BLG}=2e^2\gamma_1/[4\pi\varepsilon(\hbar v_0)^2] \sim 1$ nm$^{-1}$.

For the charged impurity scattering we 
use the ``complete-screening'' approximation.\cite{adam08}
Thus we find
$\tau(E)=\frac{4\hbar}{\pi^2}\frac{\gamma_1}{(\hbar v_0)^2}\frac{1}{n_{imp}}$,
where $n_{imp}$ is the average impurity density.
This yields
\begin{equation}\begin{split}
\sigma(E) = C_{BLG}|E|, \quad C_{BLG}=\frac{8e^2\gamma_1}{\pi^³\hbar^3v_0^2n_{imp}}.
\end{split}\end{equation}
Then the transport coefficients are
\begin{equation}\begin{split} \label{e.blgtrcoeffs}
\hat{\sigma}(\mu,T) =& C_{BLG} 2 k_B T\ln\left(2\cosh\frac{\mu}{2k_BT}\right) \\
\hat{\kappa}(\mu,T) =& 
\mathcal{L}T C_{BLG} \frac{3}{\pi^2}k_BTh(\mu/k_BT) \\
\hat{\gamma}(\mu,T) =& 
C_{BLG}2k_BT[-\frac{\mu}{k_BT}\ln(2\cosh(\frac{\mu}{2k_BT}))  \\
&+\mathrm{Li}_2(-e^{-\mu/k_BT}) - \mathrm{Li}_2(-e^{\mu/k_BT}) 
],
\end{split}\end{equation}
where $\mathcal{L}=\frac{\pi^2}{3}\frac{k_B^2}{e^2}$ is the Lorenz number.
Here and $\textrm{Li}_2$ is the dilogarithm function.\cite{nistbook}
and $h(a)=h(-a)$ is defined as
\begin{equation}\begin{split}
h(a)=\int_{-\infty}^{\infty}|x|(x-a)^2
\left[-\frac{d}{dx}\frac{1}{e^{x-a}+1}\right]dx.
\end{split}\end{equation}
This function has the limits 
$h(a)\approx \frac{\pi^2}{3}2\ln(2\cosh(a/2))$, when $|a|\gg 1$,
and $h(a)\approx 9\zeta(3)$, when $|a|\ll 1$.
Using these we see that the Wiedemann-Franz law $\kappa=\mathcal{L}T\sigma$
only applies if $|\mu|\gg k_B T$.

\subsection{Monolayer graphene} \label{s.mlgcoeffs}

For MLG the dispersion relation is $E=\pm \hbar v_0 k$, giving
a constant group velocity $v(E)=v_0=10^6$ m/s, 
and a density of states
$\mathcal{D}(E)=\frac{1}{\pi}\frac{2|E|}{(\hbar v_0)^2}$.
Then
\begin{equation}\begin{split}
\hat{n}(\mu,T) = \frac{1}{\pi}\frac{2(k_BT)^2}{(\hbar v_0)^2}
g(\mu/k_BT).
\end{split}\end{equation}
Here we have defined the function 
$g(a) = \mathrm{Li}_2(-e^{-a}) - \mathrm{Li}_2(-e^{a})=-g(-a)$,
which has the limits $g(a)\approx a|a|/2$, when $|a|\gg 1$, and
$g(a)\approx 2\ln(2)a$, when $|a|\ll 1$.
The inverse screening length is now
$q_{TF,MLG}= 4 k_F e^2/(4\pi\varepsilon \hbar v_0)$,
with $k_{F}=|\mu|/(\hbar v_0)$.

For the impurity scattering we now assume 
that the ``effective fine-structure constant'' \cite{adam07b,adam08}
of MLG is small, 
$r_s=q_{TF,MLG} / (4k_{F})=e^2/(4\pi\varepsilon\hbar v_0)\ll$ 1. 
(For SiO$_2$ $\varepsilon=4.0$, and $r_s\approx 0.5$.)
In this way we find
$\tau(E)=\frac{1}{n_{imp}}\frac{\hbar}{\pi^2}\frac{|E|}{(\hbar v_0)^2}\frac{1}{r_s^2}$.
These give
\begin{equation}\begin{split}
\sigma(E) = C_{MLG}|E|^2, \quad C_{MLG} =\frac{e^2}{\pi^3r_s^2\hbar^3v_0^2n_{imp}}.
\end{split}\end{equation}
The transport coefficients are thus
\begin{equation}\begin{split} \label{e.mlgtrcoeffs}
\hat{\sigma}(\mu,T) =& C_{MLG} 
\left[\mu^2 + \frac{\pi^2}{3}(k_BT)^2\right]  \\
\hat{\kappa}(\mu,T) =& \mathcal{L}T  C_{MLG} 
\left[\mu^2 + \frac{7\pi^2}{5}(k_BT)^2\right]  \\
\hat{\gamma}(\mu,T) =& \frac{2\pi^2}{3}C_{MLG}\mu k_BT. \\
\end{split}\end{equation}
Note again that the Wiedemann-Franz law $\kappa=\mathcal{L}T\sigma$ 
is only approximately valid in the limit $|\mu|\gg k_BT$.


\section{Low-bias resistance of $p$-$n$ junction: classical
thermal activation vs. quantum tunneling} \label{s.pnj}

In order to understand the temperature-dependence of the 
conductivity in Fig.\ \ref{f.expcomp_mlgblg} at 
$\phi_{\bgel}\ll\phi_{\bgel,min}$, we discuss some analytical results for 
the semiclassical conductance of a $p$-$n$ junction in BLG or MLG.
The existence of a $p$-$n$ junction at location $x=x_0$ means that
$\mu_{eq}-E_D(x_0)=0$.  At low temperature we linearize $E_D(x)$
around this point, such that $\mu_{eq}-E_D(x)\approx -A_1(x-x_0)$,
where $A_1=E_D'(x_0)$.  The classical linear-response conductance for
width $W$ is then $G=W(\int_{-\infty}^{\infty}\rho(x)dx)^{-1}$, where
$\rho(x)=[\hat{\sigma}(\mu_{eq}-E_D(x),T)]^{-1}$.

\subsection{BLG}

In this case $\hat{\sigma}(\mu,T)$ is given by Eq.\ (\ref{e.blgtrcoeffs}).
Since we use the parabolic-band approximation, 
the $x$ integral diverges logarithmically and
a cutoff length $L_c$ is needed, which should be on the order of $L$.
In this way, the conductance of a $p$-$n$ junction (width $W$) may be 
approximated with
\begin{equation}\begin{split} \label{e.pnblg}
G \approx  \frac{C_{BLG}A_1W}{2\ln(A_1L_c/2k_BT)}.
\end{split}\end{equation}
The temperature-dependence has a logarithmic 
singularity at $T=0$. 
This is the behavior seen
in Fig.\ \ref{f.expcomp_mlgblg}(a) at $\phi_{\bgel}\ll\phi_{\bgel,min}$.

Clearly the semiclassical result must break down at low enough temperature,
in which case some quantum-mechanical result taking into account
Zener-Klein tunneling is needed. The zero-temperature
conductance would then remain finite.
The simplest way to approximate the crossover
temperature is to use a Wenzel-Kramers-Brillouin (WKB) 
approximation in a similar fashion as
done for MLG.\cite{cheianov06,sonin09,vandecasteele10} 
Estimates of this type show that the crossover
temperature may well be on the order of room temperature.
We note that such a calculation predicts a superlinear 
current $I\propto V^{a}$ with $a=4/3$,
unlike in MLG where $a=3/2$.

\subsection{MLG}

Here $\hat{\sigma}(\mu,T)$ is found from Eq.\ (\ref{e.mlgtrcoeffs}).
The $p$-$n$ junction in MLG has a conductance 
\begin{equation}\begin{split} \label{e.pnmlg}
G = \frac{24k_BTC_{MLG}A_1W}{\pi^3}.
\end{split}\end{equation}
The linear temperature dependence is seen in
Fig.\ \ref{f.expcomp_mlgblg}(c) at $\phi_{\bgel}\ll\phi_{\bgel,min}$.
At low temperature the $p-n$ junctions completely dominate the
conductivity of the entire sample. However, again, at low enough
temperature this result breaks down. WKB estimates shows that this may
occur already close to room temperature. Thus the Boltzmann
calculations are only valid in the absence of $p$-$n$ junctions.



\begin{thebibliography}{10}
\bibitem{footnote2}%
  \BibitemOpen
  \bibinfo {note} {A simple way of taking phonons into account would be to
  subtract the power lost to them in the heat balance. The magnitudes of the
  electron-phonon power for various types of phonons have been estimated in
  Ref.\ \onlinecite{viljas10mod}. For example the power per area for optical
  phonons of energy $\Omega$ is of the form
  $P^{(op)}(x)=q^{(op)}\{n[\Omega/k_BT(x)]-n(\Omega/k_BT_0)\}$ where
  $n(z)=1/(e^z-1)$ and we assume that the phonons do not heat above $T_0$.
  Considering the remote phonons of a SiO$_2$ substrate we may use $\Omega=59$
  meV and then $q^{(op)}\sim 1$ mW/$\mu$m$^2$, which appears to be of the right
  order for reducing the slope of $\sigma(V)$. However, such a procedure cools
  long samples more effectively, making $\sigma(V)$ length dependent. For a
  proper description, the heating of the phonons and the modification of the
  transport coefficients should be accounted for.}%
  \bibAnnoteFile{Stop}{footnote2}%
\bibitem{footnote3}%
  \BibitemOpen
  \bibinfo {note} {The use of $d=210$ nm may not be justified for the long
  sample. However, the $\sigma(V=0)$ data for it could also be fitted with the
  correct $d$ and a somewhat smaller $n_{imp}$, which is another unknown
  parameter that can vary between the samples.}%
  \bibAnnoteFile{Stop}{footnote3}%
\end{thebibliography}

%

\end{document}